# Nanodiamond in tellurite glass Part I: origin of loss in nanodiamond-doped glass

Heike Ebendorff-Heidepriem,[1*] Yinlan Ruan,[1] Hong Ji,[1] Andrew D. Greentree,[2] Brant C. Gibson[2], and Tanya M. Monro[1]

[1] *ARC Centre of Excellence in Nanoscale Biophotonics, Institute of Photonics and Advanced Sensing, The University of Adelaide, Adelaide, SA 5005, Australia*
[2] *ARC Centre of Excellence in Nanoscale Biophotonics, School of Applied Sciences, RMIT University, Melbourne, VIC 3001, Australia*
[*]*heike.ebendorff@adelaide.edu.au*

**Abstract:** Tellurite glass fibers with embedded nanodiamond are attractive materials for quantum photonic applications. Reducing the loss of these fibers in the 600-800 nm wavelength range of nanodiamond fluorescence is essential to exploit the unique properties of nanodiamond in the new hybrid material. In the first part of this study, we report the effect of interaction of the tellurite glass melt with the embedded nanodiamond on the loss of the glasses. The glass fabrication conditions such as melting temperature and concentration of NDs added to the melt were found to have critical influence on the interaction. Based on this understanding, we identified promising fabrication conditions for decreasing the loss to levels required for practical applications.

©2014 Optical Society of America

# 1. Introduction

Fluorescent nanodiamonds (NDs) have emerged as important enablers of various technologies including quantum key distribution [1], quantum metrology [2], magnetometry [3,4], electrometry [5], decoherence [6] and thermometry [7]. There are now many color centers with identified room-temperature single photon emission properties [8], most of which show fluorescence in the near infra-red.

Arguably the most important of these are the nitrogen-vacancy and silicon-vacancy color centers due to their high-brightness and relatively controllable fabrication. Additionally, the nitrogen-vacancy color center shows room-temperature spin polarization and readout, which enables single-particle optically detected magnetic resonance [9], and hence the quantum sensing applications mentioned above.

For all of these applications, the efficiency with which the emitted photons are collected is critically important. When conventional confocal microscopy is used to image the emission of the $NV^-$ centers in bulk diamond, the photon collection efficiency is generally less than 10% [10]. Optical waveguides are an efficient alternative for photon collection with the added capacity to guide this emission to external detectors.

A waveguide architecture that incorporates diamond within it can provide strong coupling by incorporating the emitter within a region where the bound optical mode intensity is high. A number of architectures including pure diamond waveguides [11,12], or hybrid waveguide/fiber structures [13-21] have been explored. These hybrid approaches exploit the waveguide/fiber's evanescent field to couple the emitters to the optical mode. However they require manual placement of the NDs onto the surface of the structure [19]. Benson et al. recently reported NDs embedded in a polymer waveguide/cavity system [22] with improved photon collection efficiency but this demonstration suffered from high loss and strong background fluorescence from the polymer material.

We have developed an approach for combining NV center quantum emitters with photonic structures by embedding NDs into tellurite glass, which is then drawn into fiber [23]. This approach allows improved efficiency of the NV emitter to be coupled to a bound mode in the fiber [24]. We selected tellurite glass as host material for the following reasons. Tellurite glasses are liquid at relatively low temperatures of 600 to 700 °C, which minimizes oxidation of the NDs while mixing the NDs into the glass melt. In addition, tellurite glasses transmit light in the NV center excitation and emission wavelength range of 500-800 nm, and have a high refractive index (n=2.0), which enhances the capture of the NV emission in the fiber core [23,24].

In our previous work [23], the NDs were successfully embedded in tellurite glass using a two-step glass fabrication process. The raw materials were first melted at a high temperature, and in the second step, the temperature of the glass melt was reduced and the ND powder was added to the melt. To disperse the NDs into the melt, the melt was swirled and then kept at the lower temperature for ~15 min. NDs embedded in tellurite glass have been demonstrated to preserve single-photon emission of their NV centers [23]. Unfortunately, the fibers made from these first glasses exhibited a prohibitively high loss (>350 dB/m) at the $NV^-$ center excitation and emission wavelength range (500-800 nm), even for very low ND concentration of 9 ppm in weight added to the glass melt. The high loss prevented use of the fibers for practical applications. This result motivated investigation of the origin of loss in ND-doped tellurite glass to understand the processes between NDs and glass melt that lead to additional loss and ultimately to reduce the loss of ND-doped tellurite glass to levels enabling practical applications.

In this paper, we provide evidence that the origin of high loss in ND-doped tellurite glass is the formation of chemically reduced species such as gold nanoparticles (GNPs) and reduced tellurium species, with the formation of these species facilitated by the doping of the glass melts with ND. We show the impact of details of the fabrication conditions on the reduced species formation and the oxidation of NDs in tellurite glasses, and determine for the first time an approach to producing low-loss ND-doped tellurite glass. The use of this understanding to achieve ND-doped tellurite fibers that exhibit both low loss and functional NDs is presented in a second paper [25].

## 2. Experimental approach

### 2.1. Glass selection and fabrication

The first ND-doped tellurite glasses [23] were fabricated using Na-Zn-La tellurite glass with composition (in mol%) 73 $TeO_2$ – 20 ZnO – 5 $Na_2O$ – 2 $La_2O_3$. This glass composition is hereafter referred to as TZNL glass. We selected TZNL glass composition as prior work in our group demonstrated the suitability of this glass for low-loss tellurite fiber fabrication using the extrusion technique for the fiber preform fabrication [26]. The extrusion technique was selected to enable ultimately fabrication of microstructured optical fibers from the ND-doped glass, which are required for investigation of single photon propagation along the optical fibers or nanomagnetometry. The properties of the TZNL glass are reported in [26].

Building on this previous work on ND-doped tellurite glass and fiber fabrication [23], all glasses reported here (except E1) are TZNL glasses that were melted in gold crucibles. Glass E1 was melted in a silica crucible using a different Zn-Na-tellurite glass composition and lower glass melting temperature. Composition, melting procedure and properties of this glass and the impact of using silica crucible are described in [25]. The use of silica crucible (instead of gold crucible as for the other glasses reported here) allowed us to separate the absorption of reduced tellurium species from gold nanoparticles.

We prepared three series of TZNL glass samples using gold crucibles. (A) undoped and (B) ND-doped glasses were melted from 20-60 g batches to prepare rectangular glass blocks of approximate dimensions 15×7×30 $mm^3$ for spectroscopic measurements to investigate the impact of fabrication conditions on loss. (C) ND-doped glasses were melted from 100-150 g batches to fabricate cylindrical glass billets of 30 mm diameter and 20-30 mm height for fiber fabrication. As this paper focuses on the spectroscopic measurements, fiber fabrication details are not reported here but elsewhere [25]. The samples made from 20 g and 100-150 g batches (or glass) are referred to as small-melt and large-melt samples, respectively. The glass melting conditions for these samples are listed in Table 1 and detailed below.

Commercially sourced raw materials were used: $TeO_2$, ZnO, $Na_2CO_3$ and $La_2O_3$ with 99.99% or higher purity. The glasses were batched in a glovebox purged with dry nitrogen (≤10 ppm water).

All glasses were melted in a two-step process. First, the glass batch was melted at a temperature $T_1$ for a time $t_1$ until all raw materials were completely molten. Then the temperature was set to the doping temperature $T_2$, ND powder was added to the hot melt and the melt with ND dwelled in the furnace at $T_2$ for a time $t_2$. Finally, the ND doped melt was cast into a preheated brass mould, annealed at ~$T_g$ (315 °C for TZNL) and slowly cooled down to room temperature. For the ND-doped glasses, short $t_2$ of 10-20 min (except for C2 and E1b) was used to minimize ND oxidation while allowing sufficient time for homogenization of the ND in the glass melt. For the undoped small-melt glasses, first a 60 g batch was melted and cast at $T_1$. Then the cast glass block was cut into three pieces of ~20 g, and two of the pieces were remelted at two different $T_2$ temperatures for 60 min.

The glasses of series A, B and C (except C7) were melted in a furnace in ambient atmosphere. The glasses C7, and E1 were melted in a furnace that was attached to a glovebox. The liner inside the furnace attached to the glovebox was purged with a gas mixture composed of nitrogen (99.999% purity) and oxygen (99.9% purity). For both furnaces, we measured the temperature at the position of the crucible (hereafter referred to as 'glass temperature') as a function of the temperature using the thermocouple that controlled the furnace operation (hereafter referred to as 'furnace temperature'). By measuring the glass temperature at a range of furnace temperatures using an empty crucible, we determined the relationship between glass and furnace temperature for both furnaces. These relationships were used to predict glass temperatures for the glass melts undertaken.

Table 1. Glass fabrication conditions and properties

| Glass No. | ND content ($ppm_w$) | melt weight (g) | $T_1$ (°C) | $t_1$ (h) | $T_2$ (°C) | $t_2$ (min) | SPR peak absorption (dB/m) | total number of emitters[d] | number of NV emitters[d] | number of GNP emitters[d] |
|---|---|---|---|---|---|---|---|---|---|---|
| A1a | 0 | 60 | 900 | 4.7 | 900 | 0 | 0 | n/m | n/m | n/m |
| A1b[a] | 0 | 20 | 900[a] | 4.7[a] | 700 | 60 | 70 | n/m | n/m | n/m |
| A1c[a] | 0 | 20 | 900[a] | 4.7[a] | 650 | 60 | 350 | 40-70 | 0 | 12 |
| A2b[a] | 0 | 20 | 800[a] | 0.4[a] | 700 | 60 | 0 | n/m | n/m | n/m |
| A2c[a] | 0 | 20 | 800[a] | 0.4[a] | 650 | 60 | 0 | n/m | n/m | n/m |
| B1 | 23 | 20 | 900 | 0.8 | 750 | 20 | 70 | 2-5 | n/m | n/m |
| B2 | 23 | 20 | 900 | 0.3 | 700 | 15 | 330 | 50-80 | 1 | 5 |
| B3 | 23 | 20 | 800 | 0.5 | 750 | 15 | 0 | 0 | n/m | n/m |
| B4 | 26 | 20 | 800 | 1.0 | 700 | 15 | 50 | 9-20 | 0 | 4 |
| B5 | 101 | 20 | 800 | 0.5 | 700 | 18 | 160 | n/m | n/m | n/m |
| C2[b] | 25 | 150 | 800 | 1.0 | 700 | 30 | 200 | n/m | n/m | n/m |
| C3 | 20 | 150 | 800 | 1.0 | 700 | 12 | 130 | 170 | 11 | 10 |
| C7 | 0 | 150 | 800 | 0.6 | 700 | 11 | 0 | n/m | n/m | 10 |
| E1a[c] | 21 | 100 | 690 | 0.7 | 570 | 10 | 0 | n/m | n/m | n/m |
| E1b[c,a] | 21 | 100 | 690[a] | 0.7[a] | 610 | 30 | 0 | n/m | n/m | n/m |

[a] sample made via remelting glass at $T_2$. The $T_1$ value is the $T_1$ used for the corresponding batch-melted glass.
[b] lid used at ND doping step
[c] glass melted in silica crucible (Note other glasses melted in gold crucible.)
[d] n/m is not measured..

The ND particles used for glass doping were commercially available ND with ~45 nm diameter (NaBond). By measuring the weight of the ND powder before adding it to the melt and measuring the amount of ND powder that remained in the container used to add the ND powder, we determined the weight of ND added to the known weight of the glass melt, yielding the ND doping concentration in ppm-weight. As there was some variation how much of the ND powder remained in the container, the ND doping concentration varied up to 5 ppm from the targeted value. As detailed in Section 6, we observed oxidation (burning) of ND in the hot glass melt. Therefore, the ND doping concentration refers to the amount of NDs added to the melt, and does not specify the amount of ND that survived in the glass after the melting procedure.

As described below, we observed that the degree of ND oxidation depended on the surface-to-volume ratio of the glass melt in the crucible. For the 100 mL gold crucible used here the surface-to-volume ratio of a tellurite melt is 3.35 $cm^{-1}$ for 20 g melt, 0.84 $cm^{-1}$ for 100 g melt, and 0.59 $cm^{-1}$ for 150 g melt. Note that compared to a 150 g melt, a 100 g melt has only a 1.4 times higher ratio, whereas a 20 g melt has a 5.7 times higher ratio.

*2.2. Spectroscopic and microscopic measurements*

We used two methods for glass and fiber characterization: optical absorption spectroscopy and scanning confocal fluorescence microscopy. For absorption spectroscopy measurements, we prepared polished glass samples of 3-6 mm thickness from the glass blocks and billets

(except for C3 we used the rod that was obtained by extrusion of the glass billet, for extrusion details see [25]). Absorbance spectra in the wavelength range of 200-800 nm were measured using commercial spectrophotometer (Agilent Cary 5000). Loss spectra were obtained from the absorbance spectra by subtracting Fresnel reflection and normalization to the sample thickness. Fresnel reflection can be calculated using the refractive index of the glass measured. The refractive index of undoped TZNL glass is known [27]. As the UV edge of most ND-doped glasses is identical with that of the undoped glass, we assume that the index of the ND-doped glasses is similar to that of the undoped glass, and therefore used the index of the undoped glass to calculate Fresnel reflection for all samples considered here.

Scanning confocal fluorescence microscopy was used to detect the presence of fluorescent emitters in the samples. For the small-melt glass samples, we imaged the polished face of a glass plate. For the unstructured fiber C3, we imaged the cleaved endface of short fiber pieces (~1 cm length). An excitation power of 500 µW was used at a wavelength λ =532 nm. The confocal images were taken in a spectral window of 650-750 nm. The image size was 100 µm × 100 µm with focal depth of ~1 µm and pixel size of 300 nm. The images were taken by focusing 2-10 µm below the surface to ensure that the NDs imaged were below the tellurite surface. The number of emitters in an image was determined using software ImageJ. For some samples, the emission spectrum of selected emitters was measured in the spectral window of 550-850 nm. In addition, the presence of single-photon-emitting NV center(s) inside ND was tested using an in-house built Hanbury Brown and Twiss single photon antibunching system [28] collecting emission in the 650-750 nm wavelength range.

## 3. Mechanism of gold nanoparticle formation in tellurite glass

The first ND-doped tellurite glasses were of TZNL composition and were prepared via batch melting at 900 °C and then doping with 10-300 ppm ND at 700 °C using a gold crucible [23]. The glasses doped with high ND concentration of ~300 ppm were dark, with limited transmission in the visible. For ND doping concentration in the range 10-30ppm, we observed dichroic behavior – blue color in transmission and brown color in reflection (Figs. 1a and b).

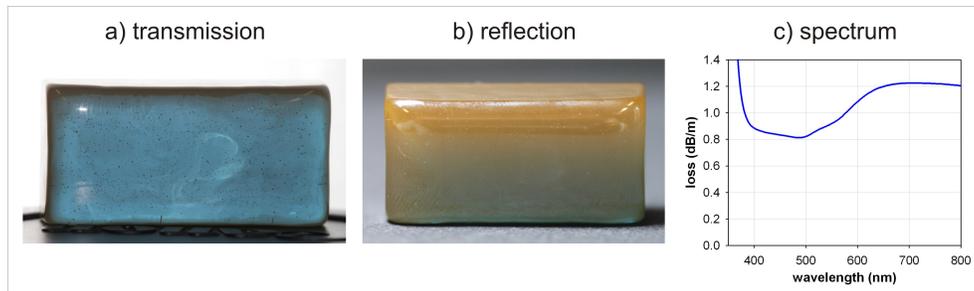

Fig. 1. ND-doped tellurite glass sample melted at 900°C and doped with ND at 700°C (made prior to the glass series considered here): Photographs taken in (a) transmission mode and (b) reflection configuration, and (c) absorbance spectrum of the sample.

To test whether the observed coloration could be caused by the absorption of NV centers in the ND powder used for the doping, we prepared a control sample by dispersing NDs in aqueous solution at room temperature using a high doping concentration of 130 ppm. Even for this high ND concentration, no absorption above the detection limit of 1 dB/m could be detected, indicating that the loss of ND-doped glasses and fibers could not be caused by the intrinsic properties of the NDs themselves.

Next, we researched the cause of dichroic behavior of glass. The same dichroic behavior was reported for high-index glass (having similar index as our tellurite glass) containing GNPs [29,30]. In addition, well-known gold-ruby silicate glasses (having lower refractive index) exhibit dichroic behavior [31] – red color in transmission and green color in reflection. The dichroic behavior of our glasses is demonstrated in the loss spectrum through high background loss and broad absorption band at ~700 nm (Fig. 1c). The high background loss is attributed to scattering of the GNPs (causing the brown color in reflection). The broad band at ~700 nm is attributed to surface plasmon resonance (SPR) absorption of GNPs (causing blue color in transmission).

The dichroic behavior, high background loss and broad absorption band >600 nm of our first ND-doped tellurite glasses resemble the dichroic behavior and absorption spectra of GNPs in heavy metal oxide glass having similarly high refractive index as our tellurite glass [29,30]. Based on this similarity in absorption and scattering behavior, we conclude that our ND-doped glasses contain GNPs, and that the GNPs are responsible for the high loss in the first ND-doped glasses made. The GNPs are formed in the glass since gold is incorporated into the glass via corrosion of the gold crucible by the hot glass melt.

It is known that gold exists in an ionic form in silicate glasses that are melted and quenched at high temperature (~1400 $^{o}$C) [31]. Such ionic gold is not detrimental to glass absorption over the 500-800 nm band, nor does it induce scattering. However, when heat treating glasses with gold ions at lower temperature (500-700 $^{o}$C), GNPs are formed [31]. The presence of GNPs in our tellurite glasses is therefore attributed to the following mechanism. During high temperature melting (900 $^{o}$C), gold from the gold crucible dissolves into the tellurite melt in ionic form. When NDs are added to the melt at lower temperature, the gold ions transform into GNPs, causing high loss and dichroic behavior of the glasses. Based on this hypothesis, we can explain the impact of fabrication conditions on the loss of undoped and ND-doped glasses as described in the next Section.

## 4. Impact of temperatures ($T_1$ and $T_2$) and melt volume on GNP formation in ND-doped tellurite glasses

To further investigate the formation of GNPs, we determined the loss spectra of small-melt glasses both undoped and doped with ~25 ppm ND and prepared using different temperatures ($T_1$, $T_2$) for the first and second step of the glass fabrication process. Figures 2a and c show that some samples exhibit characteristic SPR absorption >600nm. To separate the loss due to SPR absorption from the background loss for these spectra, we subtracted the minimum loss at ~500 nm from the loss spectra, resulting in so-called SPR loss (Figs. 2b and 2d). The peak absorption value of the SPR loss spectra (listed in Table 2) is used as a measure of the amount of GNPs in glass. Figures 2b and 2d show the SPR loss spectra relative to a spectrum without SPR absorption. Due to small sample thickness of <10 mm, only SPR absorption >20 dB/m could be determined with sufficient reproducibility (<10% variation). In addition, for thin samples, surface imperfections lead to high background loss, which can be higher than the background loss due to scattering defects in the glass volume. As it is difficult to achieve excellent surface quality for polished tellurite glass samples due to their ability to be scratched easily, the background loss is only semi-quantitatively studied here.

The undoped glasses melted at high $T_1$ = 900 $^{o}$C and low $T_2$ = 700-650 $^{o}$C exhibit both high background loss and high SPR intensity (Figs. 2c,d), indicating the formation of GNPs. The ND-doped glasses melted at high $T_1$ = 900 $^{o}$C and/or low $T_2$ = 700 $^{o}$C demonstrate same SPR absorption as for the undoped glasses (Figs. 2a,b), indicating that GNPs were also formed in the ND-doped glasses. These results confirm that temperature plays a key role in the precipitation of GNP. The impact of $T_1$ and $T_2$ on SPR absorption for both undoped and ND-doped glasses is illustrated in Figure 3. The SPR peak absorption increases with increasing $T_1$ and decreasing $T_2$, confirming that high $T_1$ and low $T_2$ facilitate GNP formation. However, for the ND-doped samples, similar SPR peak absorption is observed for ~50 $^{o}$C higher $T_2$. This result demonstrates that in the ND-doped glasses similar concentrations of GNPs are formed at higher $T_2$ compared with the undoped glasses. In other words, for the same $T_2$, a larger number of GNPs is formed in the ND-doped glasses compared with undoped glasses, indicating that ND doping serves to *enhance* the GNP formation.

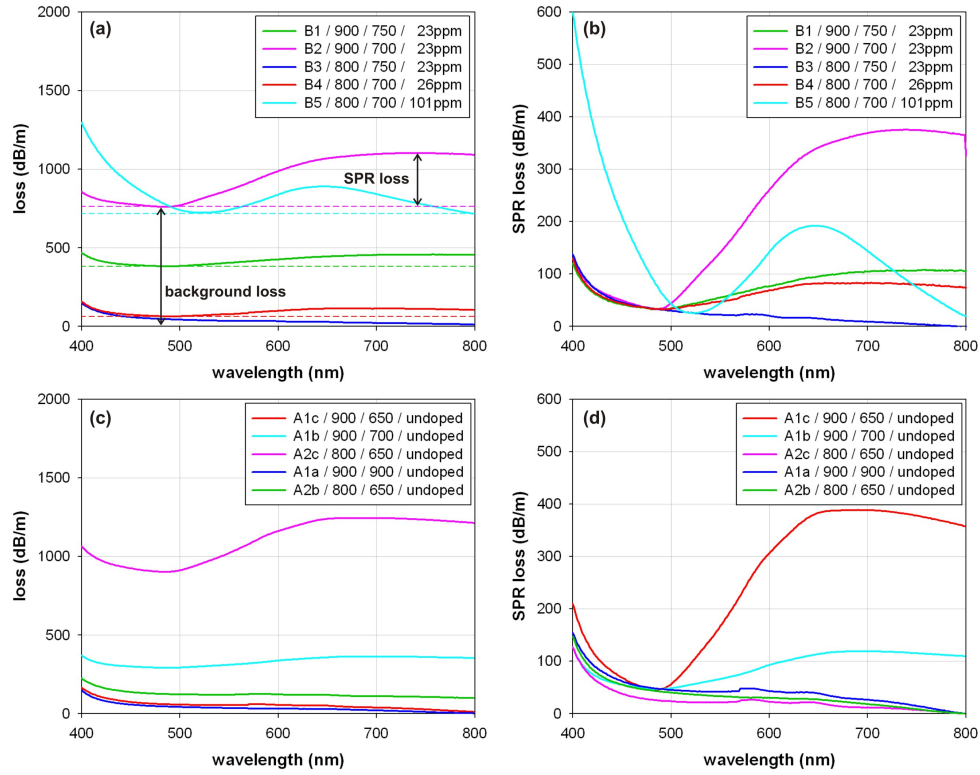

Fig. 2. Absorption behavior of small-melt glass samples that were either (a,b) doped with ~25ppm ND or (c,d) undoped. The glasses were melted using different temperatures $T_1$ and $T_2$. (a,c) show the loss spectra, while (b,d) show the SPR absorption extracted from the loss spectra.

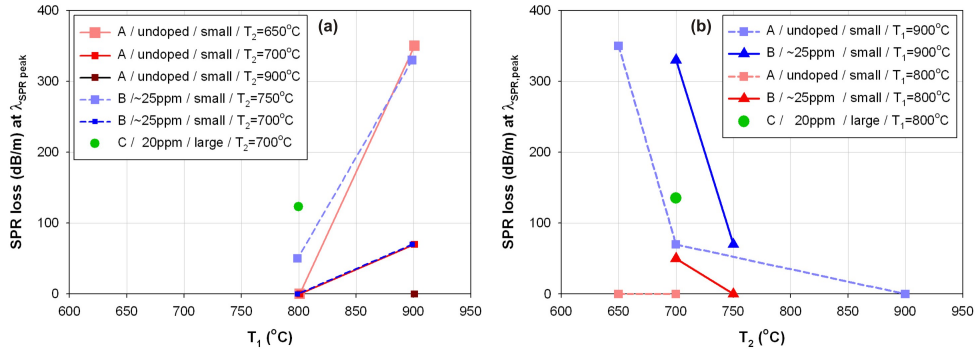

Fig. 3. SPR peak absorption of undoped and ND-doped small-melt and large-melt samples melted in gold crucible as a function of (a) $T_2$ for different $T_1$ and (b) $T_1$ for different $T_2$.

We attribute the enhancement of GNP formation by ND doping to the reducing effect of the carbon in ND. To evaluate the reducing effect of ND, we also investigated the reducing effect of graphite. Graphite was selected as it is a carbon compound like ND but has different bonding – $sp^2$ hybridized carbon atoms compared to $sp^3$ hybridized carbon atoms in diamond.

We prepared glasses melted using the same melting conditions as for B4 ($T_1$=800 °C, $T_2$=700 °C) and B2 ($T_1$=900 °C, $T_2$=700 °C) but doped with 50 ppm graphite. Note that the molar amount of 50 ppm graphite corresponds to 31 ppm ND due to the density of graphite (2.2 g/cm$^3$) being lower than that of diamond (3.52 g/cm$^3$). The graphite doped glass corresponding to B4 (low $T_1$) did not show any SPR absorption, while the graphite doped glass corresponding to B2 (high $T_1$) exhibits a similar SPR peak absorption to B2. Note that higher molar concentration of graphite was required to induce the same SPR absorption compared with ND. The test melts with graphite indicate that ND has a higher chemically reducing effect towards gold ions than graphite. We hypothesize that this is caused as follows. Graphitic (sp$^2$ carbon) is oxidized in air at lower temperature (>375 °C) than nanodiamond (sp$^3$ carbon) (>450 °C) [32]. Thus, we assume that nanodiamond reacts slower with oxygen dissolved in the melt. Due to reduced reaction rate with oxygen, nanodiamond can react with gold ions in the glass matrix, resulting in GNPs.

At $T_1$=800 °C, comparison of the large-melt sample C3 and the small-melt sample B4 with similar ND doping concentration of 20-23 ppm demonstrates that the SPR peak absorption increases with increasing melt volume (Fig. 3). This result and the GNP formation enhancement by ND are discussed further in Sections 5 and 6 (below).

## 5. Impact of ND doping concentration on GNP and reduced tellurium species formation

For small-melt TZNL samples made using the same $T_1/T_2$ temperatures, SPR absorption intensity increases with ND concentration (Fig. 4a, Table 1), confirming that ND doping enhances GNP formation.

B5 glass with high ND doping of 101 ppm and melted in a gold crucible does not only show a strong SPR peak but also a shift of the UV edge to longer wavelength (Figs. 2a,b). This is attributed to the formation of reduced tellurium species (Te$^0$ and/or Te$^{2-}$). A glass sample with very high ND doping concentration of 250 ppm is black with no transmission in the visible. We attribute this result to the formation of such a large amount of reduced tellurium species that the short wavelength edge of the glass is shifted to longer wavelengths beyond the visible range.

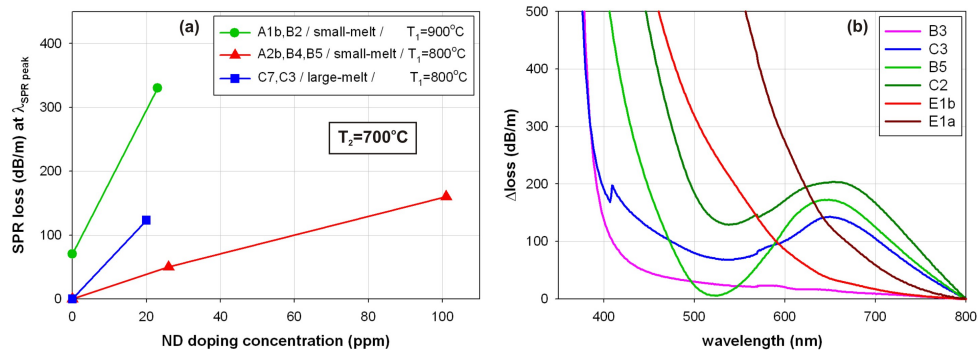

Fig. 4. (a) SPR peak absorption as a function of ND doping concentration for TZNL small-melt and large-melt samples, and (b) loss spectra determined by subtracting the minimum absorption for glass samples with different short wavelength edge position.

Generally, melts with larger ND doping concentration (>20 ppm) and reduced oxygen in the melt (due to larger melt size, use of lid in the doping step, low oxygen in atmosphere) can have dark coloration (dark green, brown, black). In Figure 4b, the absorption spectra of glasses with dark coloration melted in gold or silica crucibles (B5, C2, E1) are compared with an undoped glass without SPR absorption (B3) and a ND-doped glass with SPR peak but no dark coloration (C3). These spectra confirm that the dark coloration is caused by a shift of the

UV edge to longer wavelengths. The shift of the UV edge for E1 glasses melted in a silica crucible and thus being essentially free of gold ions highlights that the shift is not caused by GNPs but reduced tellurium species. The high ND doping concentration combined with large melt volume (i.e. reduced amount of oxygen dissolved in the melt) for E1a glass facilitated reaction of ND with Te-O bonds, leading to reduced tellurium species and consequently a long wavelength shift of the UV edge. Glass E1b was made by remelting E1a in 100% oxygen atmosphere. The shift of the UV edge back to shorter wavelengths for E1b compared to E1a indicates that during remelting reduced tellurium species were oxidized.

Considering both GNP and reduced tellurium species formation, we found that for low ND doping concentrations (<30 ppm for small melts), increasing ND doping concentration results in increased number of GNPs with no formation of reduced tellurium species. At high ND doping concentrations (≥100 ppm in small melts), increasing ND doping concentration increases the number of reduced tellurium species. The ability to melt glasses with GNPs but without reduced tellurium species demonstrates that the chemical reduction of gold ions by NDs has a higher probability than the reduction of $Te^{4+}$ by ND.

Based on this understanding of the impact of glass melting temperatures and ND doping concentration, we attribute the high loss of the first ND-doped tellurite glasses melted at high $T_1$ of 900 °C [23] to the following effects. The dark color combined with high ND doping concentration (~300 ppm) suggests that the first ND-doped tellurite billet made contained a large amount of reduced tellurium species. It is likely that the glass also contained GNPs. The higher transparency and slightly dichroic behavior of the second ND-doped tellurite glass billet prepared using low ND doping concentration (9 ppm) makes it likely that this glass contained GNPs but no reduced tellurium species.

## 6. Confocal microscopy results for small and large melts: Detection of NDs in the tellurite glasses

The previous section described the impact of fabrication conditions on the interaction of NDs with gold ions dissolved in the glass and the effect of this interaction on glass loss. This section investigates the impact of fabrication conditions on the survival of NDs added to the hot glass melt.

Diamond is known to react with oxygen at ≥650 °C, resulting in graphitization and eventually oxidation to carbon oxides. NDs already react with oxygen at lower temperatures ≥450 °C [32]. As described in Section 5, tellurite glasses need to be melted in oxygen-containing atmosphere to avoid formation of reduced tellurium species that exhibit high absorption in the visible. Therefore, NDs doped into tellurite glass melts are exposed to oxygen when dissolved in the glass melt via interaction of the glass melt with the surrounding atmosphere. The presence of oxygen combined with doping temperatures of >500 °C can destroy NDs via oxidization. We investigated whether NDs survived in the tellurite glass melts using confocal microscopy at room temperature.

Before we studied ND-doped glasses, we tested whether GNPs in tellurite glass show emission by taking confocal microscopy images of undoped tellurite samples. No emitters were found for glasses that did not show SPR absorption, whereas large number of emitters was found in the undoped glass A1c showing strongest SPR absorption. As A1c does not contain ND, the emitters found in this glass are therefore attributed to GNPs. The spectra of all 12 GNP emitters measured for A1c show the same emission band at ~650 nm (Fig. 5). Spin-coated GNPs on a glass slide were found to show an emission band at ~600nm [33] using the same excitation wavelength of 532 nm as for our measurements. The shift of the emission band of the GNPs embedded in high-index tellurite glass (~650 nm) relative to the GNPs coated on a glass slide (600 nm) is consistent with the shift of the SPR absorption peak from ~600 nm for GNPs on a glass slide [33] to ~650-700 nm for GNPs embedded in tellurite glass. GNPs in silica fiber were found to show broad emission spectrum at 600-1400 nm [34].

Compared to the spectrum of NV centers in ND powder, the GNP band in tellurite glass is situated at shorter wavelength (Fig. 5).

The finding that GNPs show emission infers that an emitter in an ND-doped sample can be either a GNP or a ND with NV center(s). However, the finding that the peak of the GNP emission band is situated at shorter wavelength (~650 nm) compared to that of the NV emission band of ND powder (~690 nm) allows distinction between these two types of emitters in our glasses. Emission spectra with peak at ~650 nm are attributed to GNPs, whereas emission spectra with peak at ~690 nm are attributed to NV centers in NDs.

In the sample B2 melted at high $T_1$, of the 6 tested emitters, 5 emitters show GNP spectrum and only 1 emitter shows NV spectrum. In the ND-doped sample B4 melted at low $T_1$, all 4 tested emitters show GNP spectrum (Fig. 5, Table 1). These results imply that most of the NDs doped to the glasses were oxidized.

Of the 27 emitters tested for C3 melted at low $T_1$ as for B4, we found 10 to have GNP spectrum and 11 to have NV spectrum and 6 to have spectral shape of both GNP and NV (Fig. 5, Table 1). The detection of GNPs in C3 is in agreement with the strong SPR peak for this glass. The presence of emitters with NV spectral shape in C3 is in contrast to the corresponding small-melt sample B2, for which we did not find any emitters with NV spectral behavior. The origin for the different behavior of the small-melt and large-melt samples is discussed below.

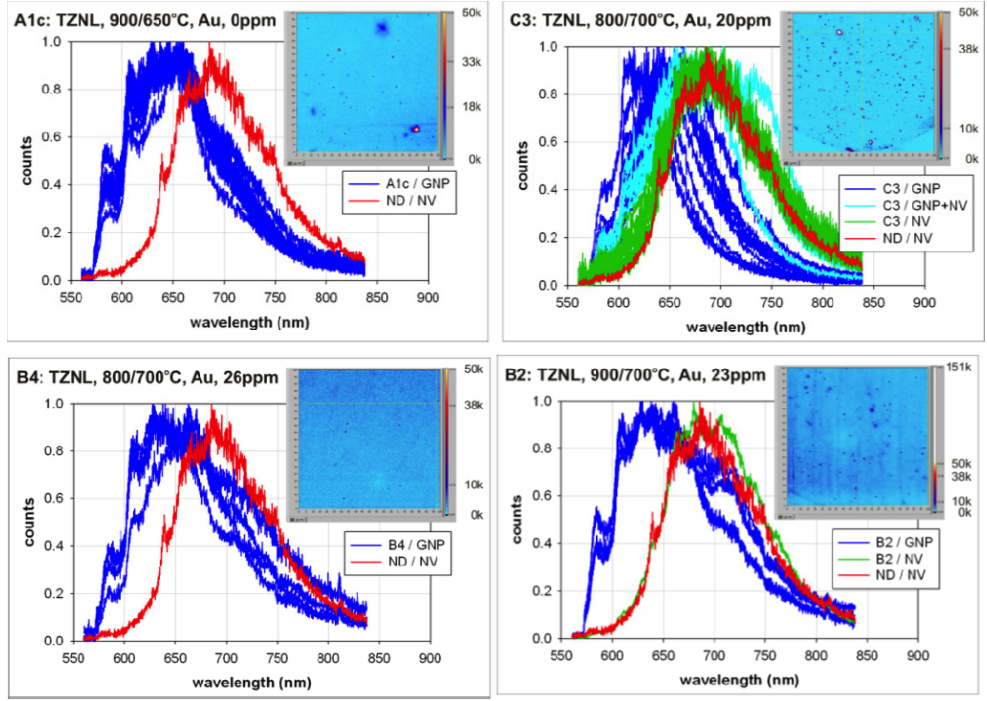

Fig. 5. Fluorescence spectra of emitters (blue, green and cyan lines) in various glass samples and of ND powder (red lines) imaged using confocal microscopy. The blue and green spectra are attributed to GNP and ND, respectively. The cyan spectra exhibit GNP and NV spectral components. The insets show the corresponding 100μm × 100μm confocal images of the samples.

Another method to detect the presence of NDs in glass is the measurement of the antibunching behavior of an emitter. Our initial work on ND-doped tellurite glasses that were melted using $T_1$=900 °C and $T_2$=700 °C [23] demonstrated antibunching behavior for emitters in the glass. This proved that at least some of the NDs doped into the glass survived and that the NV center of an ND particle embedded in the tellurite glass showed single-photon emission. We first tested the emitters in the undoped sample A1c for antibunching behavior. This sample exhibits GNPs that demonstrate strong SPR absorption and emission at shorter wavelength compared to NV centers. None of the 15 emitters tested in A1c showed antibunching, which confirms that the emitters are GNPs, which are not expected to show single-photon emission. Next, we tested ND-doped tellurite samples. As for the undoped sample, we did not observe antibunching for any of the 20 emitters tested in B2 as well as for any of the 50 emitters tested in B4. This result coincides with the observation that most of the measured emitters did not show typical NV spectral behavior but GNP spectral behavior (Fig. 5). Therefore, the absence of antibunching suggests that for the small-melt glass samples doped with 23-26 ppm ND the majority of emitters are GNPs and that almost all NDs doped into the glass melt were destroyed via oxidation within the time of 10-20min at $T_2$. The conclusion that the majority of emitters in small-melt samples are GNP agrees with the observation that the number of emitters correlates with the SPR absorption of the GNPs in the samples (Fig. 6). The occurrence of GNPs in the absence of ND at temperatures, where undoped samples do not form GNPs, indicates that the GNPs are formed as a result of ND doping and that they are stable for some time after NDs have been completely vanished due to oxidization. The larger number of gold ions at $T_1=900\ ^oC$ resulted in the formation of numerous GNPs for the same ND doping concentration compared with lower $T_1=800\ ^oC$.

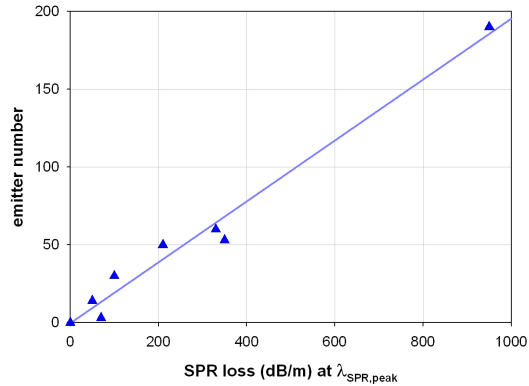

Fig. 6. Number of emitters as a function of SPR absorption at the peak wavelength for small-melt samples with different ND doping concentrations and melted using different $T_1$ and $T_2$.

In contrast to the ND-doped small-melt sample B4, the corresponding large-melt sample C3 exhibited emitters with antibunching behavior, which is consistent with the presence of NV emitters in C3 but absence of NV emitters in B2. This result indicates that ND particles survived in the large-melt sample, whereas no ND survival was observed for the small-melt sample. Larger melt volume while using the same crucible has two effects. One effect is that the glass melt takes longer time to cool down when taken out of the furnace. As a result, the NDs are exposed to higher temperature when NDs are added to the melt and during casting. The exposure to higher temperature should facilitate ND oxidation, however we found the opposite result. The other effect of large melt volume is a significantly lower surface-to-volume ratio (1/6 of the ratio of small melts, see Section 2.1), which reduces interaction of the melt with the atmosphere and thus with oxygen. The reduced oxygen in the larger melt is assumed to reduce oxidation of NDs by oxygen, leading to a larger number of NDs surviving in the quenched glass.

## 7. Conclusions

We investigated the impact of a range of fabrication conditions on the loss of ND doped tellurite glass samples and the oxidation of ND in the glass melts. GNPs and reduced tellurium species cause high loss in the visible. GNPs have strong SPR absorption at 600-700 nm in tellurite glasses, whereas reduced tellurium species cause a shift of the UV edge into the visible spectral region. Tellurite glasses that are melted in a gold crucible contain gold ions due to corrosion of the gold crucible by the hot glass melt, in particular at high batch melting temperatures in excess of $T_1 \geq 800\ ^oC$. GNPs can be formed from the dissolved gold ions depending on glass melt temperature and ND doping concentrations. For both small-melt and large-melt samples, doping of ND facilitates the formation of GNPs. Hence, for melts with sufficient amount of gold ions to form GNPs in quantities that can be detected, increasing ND doping concentration results in an increase in the GNP content. The batch melting temperature $T_1$ has a strong impact on GNP formation via increase of the amount of dissolved gold ions with increasing $T_1$. The ND doping concentration is limited to <20 ppm by the formation of reduced tellurium species at higher ND doping concentrations.

GNP emitters show shorter emission peak wavelength than NV emitters, allowing the type of emitters in our glasses to be identified.

The higher ND survival probability for large melts is attributed to smaller surface-to-volume ratio, resulting in reduced interaction of the melt with oxygen in the atmosphere, which reduces the probability of oxidation of ND by dissolved oxygen in the melt.

In conclusion, we identified the following fabrication conditions to reduce the loss in ND-doped glass and enable survival of NDs doped into a melt. The batch melting temperature needs to be as low as possible to avoid dissolution of gold from the crucible into the glass melt. Smaller surface-to-volume ratio favors survival of NDs but enhances the formation of GNPs from gold ions dissolved in the glass. The ND doping concentration is limited to ~20 ppm; higher concentration can lead to GNP and/or reduced tellurium species formation. However, as is shown in Part II of this study, for large melts with small surface-to-volume ratio, 20 ppm ND doping concentration is more than sufficient to obtain single photon propagation along practical fiber lengths.

Following this systematic investigation of the impact of the fabrication conditions on the quality of the tellurite glass and the preservation of the nanodiamond emission signature, in Part II of this study we present developments in the production of practical ND-doped optical fibers.

## Acknowledgements

This work has been supported by ARC grants (DP120100901, DP130102494, FF0883189, FT110100225, LE100100104, CE140100003). T.M. acknowledges the support of an ARC Georgina Sweet Laureate Fellowship. This work was performed in part at the OptoFab node of the Australian National Fabrication Facility utilizing Commonwealth and SA State Government funding. We wish to thank Rachel Moore and Philipp Köder (formerly at The University of Adelaide) for help with glass melting.